\documentclass[a4paper,fleqn]{cas-dc}

\usepackage[numbers]{natbib}
\usepackage{xcolor}

\def\tsc#1{\csdef{#1}{\textsc{\lowercase{#1}}\xspace}}
\tsc{WGM}
\tsc{QE}
\tsc{EP}
\tsc{PMS}
\tsc{BEC}
\tsc{DE}
\begin{document}
\let\WriteBookmarks\relax
\def\floatpagepagefraction{1}
\def\textpagefraction{.001}
\shorttitle{Context-Aware Environment Monitoring}
\shortauthors{Guilherme Rotth Zibetti et~al.}

\title [mode = title]{Context-Aware Environment Monitoring to Support LPWAN-based Battlefield Applications}
%\tnotemark[1,2]

%\tnotetext[1]{This document is the results of the research
 %  project funded by the National Science Foundation.}

%\tnotetext[2]{The second title footnote which is a longer text matter
 %  to fill through the whole text width and overflow into
 %  another line in the footnotes area of the first page.}

\author{Guilherme Rotth Zibetti}[orcid=0000-0002-4429-718X]

\ead{guilherme@ufrgs.br}
\address{Institute of Informatics}
\address{Federal University of Rio Grande do Sul, Porto Alegre, Brazil}

\author{Juliano Araujo Wickboldt}[orcid=0000-0002-7686-8370]
\ead{jwickboldt@inf.ufrgs.br}
\author{Edison Pignaton de Freitas}[orcid=0000-0003-4655-8889]
%\fnmark[3]
\ead{epfreitas@inf.ufrgs.br}

%\credit{Data curation, Writing - Original draft preparation}

%\address[2]{Sayahna Foundation, Jagathy, Trivandrum 695014, India}

%\author%
%[1,3]
%{Rishi T.}
%\cormark[2]
%\fnmark[1,3]
%\ead{rishi@stmdocs.in}
%\ead[URL]{www.stmdocs.in}

%\address[3]{STM Document Engineering Pvt Ltd., Mepukada,
 %   Malayinkil, Trivandrum 695571, India}

%\cortext[cor2]{Principal corresponding author}
%\fntext[fn1]{This is the first author footnote. but is common to third
 % author as well.}
%\fntext[fn2]{Another author footnote, this is a very long footnote and
 % it should be a really long footnote. But this footnote is not yet
 % sufficiently long enough to make two lines of footnote text.}

%\nonumnote{This note has no numbers. In this work we demonstrate $a_b$
 % the formation Y\_1 of a new type of polariton on the interface
 % between a cuprous oxide slab and a polystyrene micro-sphere placed
 % on the slab.
 % }

\begin{abstract}
The use of IoT-related technologies is growing in several areas. Applications of environmental monitoring, logistics, smart cities are examples of applications that benefit from advances in IoT. In the military context, IoT applications can support the decision-making process by delivering information collected directly from the battlefield to Command, Control, Communications, Computers, Intelligence, Surveillance and Reconnaissance (C4ISR) systems. Taking the benefit of the installed IoT network in the battlefield, the use of the data collected by the IoT nodes is a way to improve resiliency and increase the survivability of networks, as well as to optimize the use of available resources. Towards improving the communication network present on the battlefield, this work presents a context-aware~environmental~monitoring~system that uses real-time battlefield information to increase military networks' resilience and survivability. The proposed approach is validated by a proof-of-concept experiment. The obtained results show that the implementation of this system can improve the communication process even when the network is exposed to unfavorable climatic factors.
\end{abstract}


\maketitle
\section{Introduction}
\label{sec:1}

The discussion about the convergence of the real and cybernetic world is not recent, and one of the paradigms with the most significant impact in this context is the Internet of Things (IoT)~\cite{conti2012looking}\cite{borgia2014internet}. Nowadays, IoT applications are employed to solve problems in several fields such as agriculture~\cite{dobrescu2019context}, smart cities~\cite{kamienski2017application}, and C4ISR battlefield systems~\cite{zheng2015leveraging}, which brought the concept of Internet of Battle Things (IoBT) \cite{7756279}. In particular, the military context involves mission-critical applications with real-time requirements \cite{Zac17DTN}, so it is necessary to rely on resilient and reliable networks to support these applications~\cite{plesse2005olsr} \cite{7756279}.

Among the communication technologies developed to implement IoT applications, it is possible to highlight Low-Power Wide-Area Networks~(LPWAN). Attributes, such as low power consumption, low data rate, low implementation cost, and long signal range, drive the efforts to develop these technologies~\cite{patel2017experimental}. Although LPWAN technologies have not been developed to support military applications, recent research efforts seek to adapt their use to solve problems in this context~\cite{singh2017semantic,jalaian2018evaluating}. Like other widespread wireless networking technologies, LPWANs  are exposed to interference-related problems due to climatic and environmental factors. High temperatures \cite{boano2018impact}, rain \cite{boano2009low}, and vegetation \cite{marfievici2013environmental} are factors that can directly affect the performance and availability of communications.

Boano~\textit{et~al.}~\cite{boano2018impact} conducted a study using Long Range (LoRa) -- one of the most widespread LPWAN technologies -- showing that exposure of transmission and reception devices to high temperatures jeopardizes network reliability. When exposed to high temperatures, devices present attenuation in the received signal strength (RSS) and reduction in packet delivery rate (PDR) and may, in extreme cases, become subject to total rupture of the communication link. Traditional monitoring systems commonly use metrics, such as signal noise ratio (SNR), RSS, and PDR, collected directly from network devices as communication quality indicators to improve network performance and reliability. Such metrics are undoubtedly important, but they may still not provide enough information to anticipate or explain performance degradation or communication link failures.

Some strategies and frameworks developed for building resilient networks discuss the factors that affect the correct behavior of networked systems~\cite{sterbenz2010resilience,cetinkaya2013taxonomy}. External environmental factors can significantly impact on network operations and, therefore, must be considered in the design of dependable communications through context-aware monitoring mechanisms~\cite{hutchison2018architecture}. Context-aware systems paradigm refers to a system's ability to use information collected from its surrounding physical or virtual context to develop mechanisms to respond to events~\cite{schilit1994context}. In military networks, exposure of communication devices to climatic and environmental factors is usual. Therefore, monitoring context information, which affects communication performance and availability, helps to explain, and possibly to anticipate, network behavior. 

This paper proposes a context-aware monitoring system to improve the resiliency and increase the survivability of military tactical networks. A modern battlefield scenario already envisions a network of sensors to support applications such as surveillance, reconnaissance, and logistics. This proposal can either rely on these sensors or suggest installing a few more in the battlefield objects to collect relevant information, which allows the prediction of a possible degradation in network performance or link unavailability.
%\newpage
The main contributions of this work are the following:
\begin{enumerate}
\item The proposal of a context-aware monitoring system that -- unlike current proposals found on the literature -- uses information collected from the physical environment to improve battlefield overall communication efficiency in terms of PDR;
\item The proposal of two different approaches to dynamically adapt network configuration parameters based on environmental factors. In contrast with fixed configurations, both proposed approaches result in higher packet reception by the system and lower packet emission by the objects in the battlefield; 

\item The proposed system was designed to be incorporated into military tactical networks with minimal impact on the existing infrastructure and communication architecture; 
\item The proposed system is technology agnostic, nevertheless insights on the feasibility of implementation are provided based on a state-of-the-art open source IoT framework and related technologies.

\end{enumerate}

The remainder of this article is organized as follows. Session \ref{sec:2} reviews related studies in the field of resilience and survivability of networks, environmental impact on wireless communication technologies, and systems that use the concept of context-awareness to adapt to the context. In Session \ref{sec:scenario}, a battlefield scenario is introduced, and in Session \ref{sec:architecture} the proposed solution is presented. Session \ref{sec:implementation} presents insights that demonstrate the feasibility of implementation. Session \ref{sec:4} presents a proof-of-concept experiment of the proposed solution on a simulated environment and a discussion of the obtained results. Finally, Session \ref{sec:5} presents conclusions and future work.

\section{Related Work}
\label{sec:2}

Creating frameworks and guides with best practices is present in several research areas; it is no different in the context of resilient networks. Towards resilient networks, Sterbenz~\textit{et al.}~\cite{sterbenz2013evaluation} proposed a framework consisting of strategies, metrics, and techniques to evaluate and quantify the resilience of the architecture of an already existing and a proposed network. In this framework, one of the prerequisites is context-awareness. The authors state that, for a given network to be resilient, it is necessary to monitor the channel's conditions, the link-state, and other events that may impact its correct functioning.

As aforementioned, a context-aware system can adapt in response to external events that affect its functioning. However, for this to happen, it is necessary to monitor the surrounding environment. Some works present techniques to collect, process, and make useful the data collected from the context to support other systems. In a study developed by Preeja and Krishnamoorthy~\cite{pradeep2019mom}, the authors highlight relevant points in constructing a context-aware system, such as context modeling and organization, and the use of a middleware to simplify the development considering the heterogeneity of technologies. In a recent work developed by Pradeep~\textit{et al.}~\cite{pradeep2021leveraging}, the authors proposed a generic context model and a context organization method for IoT environments. Context modeling is related to the syntax used to represent the raw data, while the organization provides the semantics for this data and its relationships. Therefore, a system must be modeled in a way that is generic enough to support a heterogeneous environment and well organized so that the collected information makes sense for the systems that need it.

Besides gathering and processing context information, to actually improve the communication process of networks, many authors rely on the adjustment of configuration parameters. Modifying the network configuration parameters to increase the signal strength is already explored in several studies. The most commonly used parameter is transmission power control using different approaches. The first approach is at the network level \cite{park2002quantitative,narayanaswamy2002power}, which is based on applying the same transmission power to all nodes in the network. The second approach is at the node level \cite{kubisch2003distributed} by applying different transmission power to each node in the network. The third approach is based on neighborhood relation \cite{xue2004number}, in which  nodes use a different transmission power for each neighbor. A fourth approach can apply different transmission power at a packet-level \cite{lin2016atpc}. A common limitation \- among all those approaches is that the network configuration parameters were adjusted using only the information collected from the network itself, such as Received Signal Strength Indicator (RSSI) and Link Quality Indicator (LQI). While this information is useful, currently, other factors can be considered to determine network configuration parameters. In this work, we aim to explore, in addition to the commonly used quality indicators, additional information collected from the physical environment where the nodes are installed.

The work developed by Boano~\textit{et~al.}~\cite{boano2018impact} presents the impact of temperature variation, a possible climatic factor, on LoRa communication links. The results show that higher temperatures compromise the network operation and, in extreme cases, cause total link rupture. In the work developed by Luomala and Hakala~\cite{luomala2015effects}, the authors present the climatic effects of temperature and humidity on communication links using ZigBee wireless networking technology. The results of this experiment show the impact of these factors on communication links. The authors used the RSSI as a metric, showing its oscillation according to the climatic variations observed from the external environment during the experiment. In both studies by Boano \textit{et al.}~\cite{boano2018impact} and by Luomala and Hakala~\cite{luomala2015effects} modifications in configuration parameters were explored to assert the resilience and survivability of these networks when exposed to climatic factors. These studies help to emphasize how critical the monitoring of external factors is to increase network resiliency, although they are quite theoretical and do not integrate into an actual network framework or solution. 

\sloppy Towards the design of context-awareness systems with the ability to adapt to the external environmental factors, Dobrescu \textit{et al.}~\cite{dobrescu2019context} suggests the implementation of a context-aware system for the precision agriculture environments. The authors' proposal aims to integrate three emerging technologies (IoT, cloud computing, and context-awareness) to provide an irrigation system and monitor parameters related to soil properties. In this system, the environmental changes collected by the sensors determine how the system will manage water, nutrients, and pesticides. In another work, developed by Kamienski~\textit{et al.}~\cite{kamienski2017application}, the authors propose the use of IoT-related technologies to improve the process of energy consumption in public buildings and universities. In this work, the authors suggest using presence sensors and smart plugs to control room lights and monitor the energy consumption of buildings. These are examples of context-aware systems that use information collected from the surrounding physical environment to determine their behavior. Although both works use information gathered from the context, neither uses it to determine its functioning, but rather the functioning of the application they are serving.

\sloppy Another example that uses information collected from the context to define what action to take is the work proposed by Rak~\cite{rak2016new}. The author proposes adapting the topology of a wireless mesh network using the information on predicted attenuation of links based on radar measurements. More specifically, the author uses real echo rain maps to calculate attenuation in communication links and modify the network topology to improve network throughput during rainstorm periods. Differently from Dobrescu's~\cite{dobrescu2019context} context-aware precision agriculture system and the intelligent buildings of Kamienski~\textit{et al.}~\cite{kamienski2017application}, in Rak's proposal, a third-party system provides the weather maps that serve as context information. In contrast, the other systems collected context information through sensors to decide the system's action.

\sloppy In the context of military networks, it is possible to highlight three works that implement context-awareness tackling different issues. The first work developed by Papakostas~\textit{et~al.} \cite{papakostas2018energy} proposes an algorithm for building a backbone in military networks with energy-awareness to deal with power efficiency in the construction of the network topology. In the second work, Poularakis~\textit{et al.}~\cite{poularakis2019flexible} developed a system that uses the concept of Software Defined Network (SDN) to address changes in the network topology from nodes distributed across the battlefield. The third work to be highlighted was the work developed by Mishra~\textit{et~al.}~\cite{mishra2017context}. The authors propose a context-oriented proactive decision support framework that aims to accelerate the decision-making process. The developed system uses data related to the mission, environment, assets, threats, time, workload, and other factors related to the decision-making process. Based on the acquired data, the system suggests to the decision-maker some possible decisions that could be taken to solve a given problem. In addition to the works already mentioned, other studies also address the use of context-aware applications to support decision-making on the battlefield~\cite{lin2019emotion,castiglione2017context,moon2010context}. However, as in the works developed by Dobrescu \textit{et al.}~\cite{dobrescu2019context},  and Kamienski~\textit{et al.}~\cite{kamienski2017application}, the aforementioned studies in the military context use the data context to determine the application behavior. A step ahead of work in this direction is presented in \cite{8808161}, which presents an approach that comabines SDN and Information-Centric Networking (ICN) to support context awareness in IoBT systems.

The difference of the proposal presented in this current paper from the works mentioned above is that, although some of them used information collected from the physical environment, this information determined the application's behavior and not the network's behavior. In contrast, in the work developed by Rak~\cite{rak2016new}, the information that determined the system's behavior was not collected directly from the system context, but provided by a third-party system. Relying on an external entity to provide information relevant to the context of the system can cause problems due to the fact that it may not meet the time constraints of critical environments such as battlefield networks. Among the studies in the military context, it is possible highlight the works developed by Papakostas~\textit{et al.}~\cite{papakostas2018energy} and Poularakis~\textit{et~al.}~\cite{poularakis2019flexible}~,~which propose solutions for context-aware networking. However, none of these works consider the effect of environmental factors on the military networks. Other works, like the developed by Mishra~\textit{et~al.}~\cite{mishra2017context} and Leal~\textit{et~al.}~\cite{8808161} in the military context, consider context-aware applications to assist in the decision-making process of Command~and~Control~(C2)~systems, leaving out of scope the network management itself.

\section{Application Scenario}
\label{sec:scenario}

\begin{figure*}
	\centering
		\includegraphics[width=.6\linewidth]{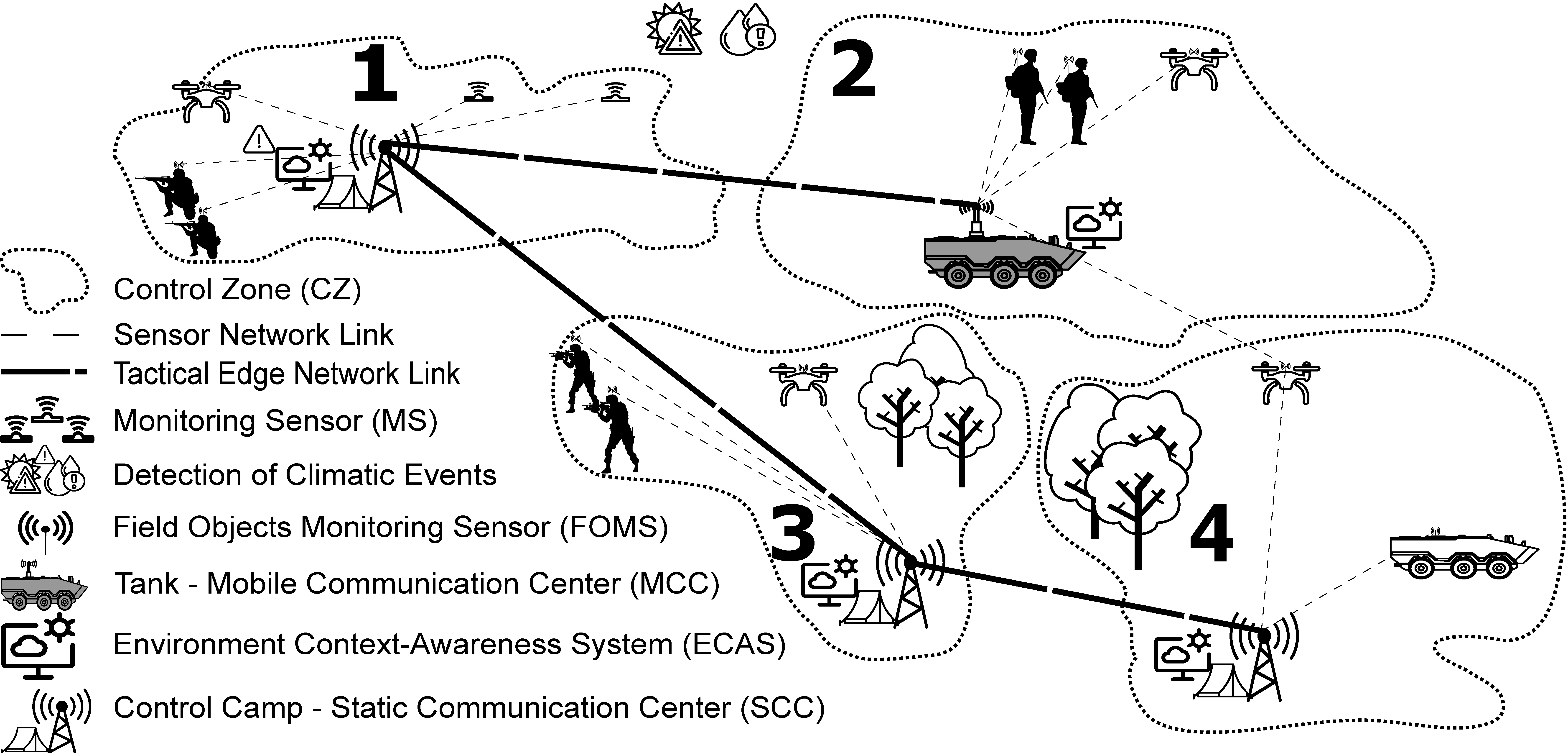}
	\caption{Battlefield Scenario.}
	\label{Battlefield_Scenario}
\end{figure*}

A fictional scenario was created containing some of the objects usually present in the battlefield to make the proposal clearer. \textbf{Figure~\ref{Battlefield_Scenario}} presents this battlefield scenario divided into Control Zones (CZ), identified by numbers from one to four. Each CZ has a Communication Center (CC), that can be Static (SCC) or Mobile (MCC), represented by control camps and a combat vehicle with a high-power antenna attached to the top (located at CZ-2). The connections between the control camps and the CZ-2 combat vehicle represent a Tactical Edge Network (TEN).

In addition to the TEN, the battlefield has a Sensor Network (SN). This network is responsible for transmitting the data from the sensors spread across the CZs to the Environment Context-Aware System (ECAS) servers located in CC of each CZ. The data traffic of the TEN has no relation to the data traffic carried out by the SN. The sensors distributed on the battlefield can communicate with any CC within range to deliver data collected from the battlefield. For instance, it is possible to observe this type of communication in CZ-4, whose flying drone has a connection with the CCs of zones two and four. Therefore, it can deliver the collected data to both CCs.

There are two types of sensors distributed on the battlefield, the Monitoring Sensor (MS) and the Field Object Monitoring Sensor (FOMS). MSs are distributed on the battlefield collecting information from a specific location, whereas FOMSs are installed on objects moving across the battlefield. Both sensors can collect temperature and humidity information. However, FOMSs can collect additional information such as vital signs, location, weaponry, and so on, from the objects to which they are attached.

The ability of MSs and FOMSs to communicate with any server makes ECAS a distributed system. Besides receiving information from multiple MS and FOMS, the ECASs servers distributed across the battlefield also share the information collected in their respective CZ through the TEN. Sharing the information collected in each CZ increases the coverage area monitored by ECAS. Therefore, this information can support decision-making and mitigate problems in any CZs that have a CC.

In \textbf{Figure~\ref{Battlefield_Scenario}}, there are two icons at the top to exemplify the system's operation, which represent the detection of climatic events related to temperature and humidity. The sensor closest to these events is the MS located in the upper right corner of the CZ-1. This sensor will notify the ECAS server of its respective CZ that it will process the received information and perform the necessary actions in response to these events. After processing the received data, the ECAS server of the CZ-1 will share the processed data with the other ECAS servers, which may need the data. After receiving the data, the other ECAS servers will process the received information to verify whether to perform an action in their respective CZ. Thus, the environment covered by the ECAS can adapt to the climatic conditions using as inputs the data collected from the battlefield through the MS and FOMS sensors.

\section{Proposed System Architecture}
\label{sec:architecture}
\begin{figure*}
	\centering
		\includegraphics[width=.7\linewidth]{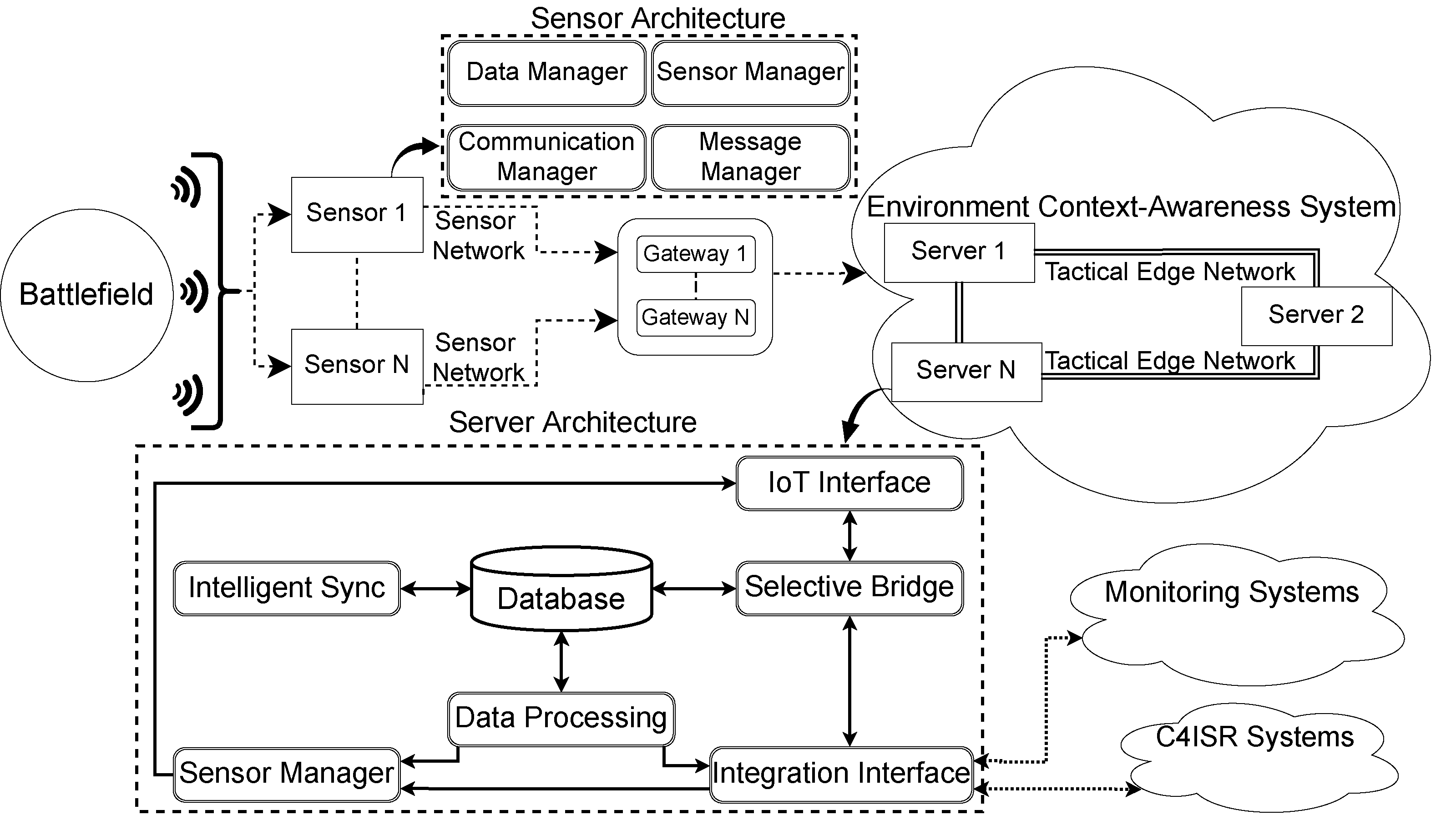}
	\caption{System Architecture.}
	\label{System_Architecture}
\end{figure*}

\textbf{Figure~\ref{System_Architecture}} illustrates the system architecture. The circle, named battlefield, refers to the monitored environment. The sensors (1 to Sensor N) represent the MSs and FOMSs distributed across the battlefield. The box with Gateway 1 and Gateway N represents the Gateways distributed by the battlefield, being a Gateway for each CZ. As it is a system that acts in a distributed way, it is necessary to have a Gateway for each ECAS server distributed by the battlefield. The dashed box on the top of the figure highlights the internal components of each sensor (i.e., the sensor architecture). 

ECAS is composed of servers that act independently and collaboratively. The distribution of these servers across the battlefield is in the same number of the existing CZ for a given military operation. That is, for each CZ, there is a Gateway and an ECAS server to handle the data collected locally. The collaborative network of servers is represented in \textbf{Figure~\ref{System_Architecture}} within the cloud shape, indicating that the communication among servers flows over the TEN. In the bottom part of \textbf{Figure~\ref{System_Architecture}} there is a highlight with the internal components of each server (i.e., the server architecture).

The following flow describes how the data collected on the battlefield is transmitted, processed, and shared. The SN is responsible for transmitting the data collected by the sensors on the battlefield to the ECAS servers distributed across the CZs. When receiving the collected data, an ECAS server must process and share it with the other servers that make part of the system. This sharing takes place through the TEN. Therefore, the SN serves exclusively for sensory data traffic and reconfiguration of sensors distributed across the battlefield. The detailing of the modules that compose the ECAS servers and sensors are presented in the following sub-sessions.

\subsection{Sensor Architecture}
\label{sec:3.3}
In \textbf{Figure~\ref{System_Architecture}}, the dashed box indicated by an arrow on Sensor 1 represents the internal architecture of each sensor. Each sensor has four modules that determine its operation. The modules are Sensor Manager, Message  Manager, Communication Manager, and Data Manager.

The Sensor Manager module is responsible for overseeing the operating mode of the sensor as a whole. Each MS and FOMS sensor has the ability to read environment data (e.g., temperature, humidity, location, vital signs) and to send this data over a SN to a gateway. The role of managing the process of collecting, storing, sending, and receiving data, which takes place through the components of each sensor, is performed by the Sensor Manager module. This module manages the frequency of data collection and local storage, as well as the frequency of transmission (in bulk). This process follows two basic criteria, which are defined as Periodic and Threshold-based. Periodic criteria means that configurable time intervals will guide the collection, local storage, and transmission of sensored data. For example, under normal conditions, temperature and humidity could be measured every few seconds, but transmitted only once per minute to save resources. Threshold-based criterion intends to reduce the frequency of data collection and transmission under unsafe operating conditions. For instance, when the measured temperature reaches a harmful threshold, it can be set for immediate transmission instead of waiting until the next periodic cycle.

The Message Manager module manages the format of the message sent to the ECAS server. Two types of messages were created: Periodic messages and Trigger messages. Periodic messages are related to the Periodic criteria, while Trigger messages are related to the Threshold-based, both defined by the Sensor Manager module. Suppose the data for a region monitored by a particular MS/FOMS sensor is outside safe operational limits. In this case, this sensor should classify the data contained in the message as urgent and send it immediately. On the other hand, if the data is within safe operational limits, the sensor can store this data locally and wait for an opportune moment for sending it, determined by the periodic criterion. Differentiating messages is a way to prioritize the communication and processing resources of the system for the most critical messages.

Defined the frequency of data collection and transmission, the types of messages, it remains to specify how to present the data to the application. The Data Manager module performs this function. ECAS aims to increase the resilience of networks as well as support the applications of C4ISR systems. Therefore, the sensor that collects and sends relevant data to ECAS does the same for C4ISR systems. The Periodic criterion determines that all sensors present in an MS/FOMS read and send the collected data to the ECAS server at a specific time. This collection criterion can be useful when the sensor is in a region where operational limits are safe; however, this is not always the case. When a sensor is in a region whose collected data is outside a safe operating threshold; the messages must contain only relevant data (i.e., data exceeding the threshold). This approach reduces the overhead of the collection and transmission process by prioritizing processing and communication resources for the most relevant data. For example, a sensor in a specific region detects that the local humidity and temperature are above the specified threshold. The Data Manager module gathers the collected data, assembles a package with the relevant data for this specific case (e.g., sensor id, collection timestamp, temperature, humidity, GPS location) and delivers this package to the Message Manager module. The Message Manager module classifies the message as Trigger and delivers it to the Communication Manager module, which sends it over the gateway to the ECAS server immediately.

The Communication Manager module is responsible for managing the communication resources of a MS/FOMS. The Sensor Manager module specifies the criteria for collecting and sending data to the ECAS server. Therefore, data collection and transmission work asynchronously, allowing the Communication Manager to set communication resources into stand-by or power-saving mode when they are not needed. The Message Manager module defines two types of messages: Periodic and Trigger, and each type have specific requirements to use the communication resources. Periodic messages must be delivered to the ECAS server as soon as possible, but may wait longer to access the medium as their delay will not significantly impact other systems. A message of type Trigger, on the other hand, contains sensitive data (i.e., data directly related to the good functioning of the system); therefore, these messages must take priority in accessing the medium and guaranteeing earliest delivery. Therefore, the Communication Manager module manages the communication resources to fulfill the message's requirements to be transmitted.

\subsection{ECAS Server Architecture}
\label{sec:3.4}
Represented in the dashed box pointed out by Server N in \textbf{Figure~\ref{System_Architecture}} is the ECAS server's architecture. Each ECAS server has the same architecture, consisting of six processing modules and one database. 

The first module of the ECAS server architecture to be discussed is the IoT Interface. IoT communication technologies provide specific protocols to meet the needs of these networks. This module is responsible for interpreting the data received from the sensors and translating them into a format consumed by ECAS and other systems that use this data. In addition to interpreting and translating sensor data, this module does the reverse process for messages sent from servers to sensors. Therefore, the IoT Interface module is the communication interface between IoT technologies and ECAS servers. The function performed by this module simplifies the implementation of heterogeneous IoT technologies on the sensor side, as it centralizes the communication process between different technologies in a single module.

ECAS uses data collected by sensors to increase the reliability of the network. In addition to data collected for ECAS, the sensors collect data relevant to C4ISR and other systems. Therefore, after the IoT Interface receives and interprets the data coming from sensors, it is necessary to forward the collected data to the respective systems of interest. The Selective Bridge module performs this function. It selectively distributes data received from the IoT Interface to be either stored and consumed locally by ECAS, as well as forwarded to external systems. This module works like a bridge between the IoT network and C4ISR or other systems, introducing only minimal overhead as data travels through ECAS. As depicted by the arrows in the ECAS Server Architecture of \textbf{Figure~\ref{System_Architecture}}, the data coming from the IoT Interface passes through the Selective Bridge module. This module then separates data that is useful to ECAS, which needs to be stored locally it in the Database, and forwards the rest to the respective systems of interest.

Once the data has been collected and separated, it is necessary to forward the data that is not useful to ECAS and process the ones that are. Before doing this processing, ECAS needs to have information from other systems available to combine with the data collected from the sensors. Information such as RSSI, SNR, and PDR are useful for ECAS to perform its role, and the providers of this information are network monitoring systems. The Integration Interface module aims to provide a single communicate interface \break among network monitoring systems, C4ISR systems, and ECAS servers, as represented in \textbf{Figure~\ref{System_Architecture}} by the arrows connecting the Integration Interface module and the clouds that represent external systems. The first function of the Integration Interface is to distribute the data received from the Selective Bridge module to the respective systems of interest and to search other monitoring systems or C4ISR for information relevant to ECAS. The second function of this module is to generate alarms for other systems (i.e., when ECAS detects an anomaly, it can generate alarms for other systems to assist in the decision-making process C4ISR systems). Finally, the Integration Interface can reconfigure the parameters of the sensors of interest of the C4ISR systems (e.g., modify thresholds of vital signs, weapons, etc.), as represented in \textbf{Figure~\ref{System_Architecture}} by the arrow that goes from the Integration Interface module to the Sensor Manager module (discussed later). This last function enables ECAS to serve as a central point for dynamic reconfiguration of parameters in all sensors in the battlefield, regardless of whether that parameter is relevant to ECAS or external systems.

The three modules described so far are mainly focused on gathering information, either from the sensors in the IoT network or from external sources, and selectively storing the gathered information in a local Database. The Data Processing module is responsible for querying the database for context information received from the IoT network (periodic and trigger messages), combining this data with information from other monitoring systems (e.g., link quality indicators), and compute potential actions that could be taken to adjust the network according to the currently assessed situation. Some possible outputs of this processing are: generating alarms for C4ISR systems (through the Integration Interface), reconfiguration of the parameters of one or a set of sensors (through the Sensor Manager), updating the quality indicators in the Database, as well as storing the processed data for later use as a dataset in the process of reasoning, which aims to anticipate the behavior of the network. 

As mentioned above, two types of sensors may be collecting data in the field, called MS and FOMS. An MS/FOMS collects data (e.g., temperature, humidity, vital signs, \break weaponry) for the ECAS and other C4ISR systems. Each system is responsible for determining the thresholds and time intervals for reading measurements. The Sensor Manager module aims at abstracting the differences between possibly heterogeneous MS/FOMS in modifying the configuration parameters. For example, if a C4ISR system needs to modify the critical threshold values of a parameter such as vital signs, this system communicates the parameters through the Integration Interface to the Sensor Manager module,\break which assembles and sends a package in the format expected by the sensor. The same process applies to ECAS itself. Suppose that a given context situation makes the Data Processing module compute a decision that requires changing the sensor's temperature or humidity thresholds. In this case, the Data Processing module instructs the Sensor Manager module to set the new configuration parameters in the sensors. %Like the Selective Interface, the Sensor Manager aims to reduce the overhead in the sensor reconfiguration process. 
The Sensor Manager, in turn, relies on the IoT Interface to communicate with the sensors in the battlefield (represented in Figure \ref{System_Architecture} by the arrow that connects the Sensor Manager and IoT Interface modules).

After collecting, distributing, processing, and storing locally the outputs of the previously described modules inside an ECAS server instance, there is still the need to verify if these outputs can be useful for other ECAS servers distributed on the battlefield. The output generated by a specific server from one CZ might be useful for servers from another CZs, but it might not be either. The function of checking which ECAS server may need each piece of data, and sharing this data is performed by the Intelligent Sync module. As already mentioned, the ECAS system relies on a set of servers distributed throughout the battlefield collecting, processing, and performing actions on the devices of their respective CZs. Due to the scarcity of resources available on the battlefield, misusing these resources can compromise network reliability. Therefore, the task of synchronizing data across servers on the battlefield should not overwhelm available network resources. The Intelligent Sync module verifies the outputs generated by the Data Processing module and decides based on context information, such as the location data of other ECAS servers and sensors, whether information should be shared or not. If the generated output is considered useful to another CZ, Intelligent Sync sends it over the Tactical Edge Network to this specific server. For example, the server located at CZ-1 may generate an output pertinent to the server located at CZ-2 (e.g., rain detected at CZ-1 and is moving towards CZ-2). Since the information related to this event is pertinent that the server located in CZ-2, this information should be sent to it as soon as possible. Nevertheless, there is no need to forward the same data to the servers in CZs 3 and 4 since the phenomenon is not likely to affect those.

\section{Implementation Feasibility}
\label{sec:implementation}

As previously explained, the system proposed in this work aims to increase the reliability and survivability of networks exposed to harsh environmental and climatic factors on the battlefield. %In the previous subsections, we have introduced a set of conceptual components that materialize our approach.
To illustrate the feasibility of the proposed system architecture, this session presents possible implementation choices for some of the components presented in \textbf{Figure~\ref{System_Architecture}}.

\sloppy FIWARE is an example of framework that aims to accelerate the development of IoT-related smart solutions~\cite{fiware_About}, which could be used to handle both data collection and issuing commands between sensors and ECAS servers. This framework provides a Context Broker (CB) component whose function is to manage context information so that other systems can consume it~\cite{fiware_Dev}. The data processed by the CB can come from several sources, including external systems or battlefield sensors. Through an IoT Agent component, the CB can collect context information, process it and make it available to ECAS. The CB can manage data collected from the context and send commands from the ECAS to sensors on the battlefield through the IoT Agent. There are ready-to-use implementations of IoT Agents based on JSON and Ultralight encoding, transporting data over AMQP, HTTP, and MQTT~\cite{fiware_IoT-Agents}. All interactions between the CB, IoT Agents, and ECAS components can take place through an open and standardized RESTful API called Next Generation Service Interface (NGSI)~\cite{fiware_Dev}.

FIWARE's NGSI API defines a data model and two interfaces for the process of exchanging context data. The main elements of the NGSI data model are entities, attributes, and metadata. An entity in FIWARE represents an object, which can be physical or virtual. This entity has attributes and metadata related to it \cite{fiware_NGSI}. Considering ECAS elements with FIWARE, an entity can represent an MS sensor, a FOMS sensor, a CZ, a CC. This entity has a unique ID and a Type (e.g., ID: Sensor-01, Type: MS; or Zone-01, Type: Control Zone). As attributes, a sensor entity could have Local Temperature, Local Humidity, Location. A CZ entity could have the location, vegetation as attributes. FIWARE entities can have relationships with each other. Therefore, it is possible to state that Sensor-01, which is of the MS type, is located in Zone-01 (i.e., it is related to Zone-01). Thus, if Sensor-01 collects and sends data, the ECAS server will know that Sensor-01 is in CZ-01.

On top of the FIWARE framework it is possible to implement several components from ECAS architecture. For example, an IoT Agent receives data collected by sensors through a Gateway and transmit commands from ECAS servers to sensors through a Gateway. Sensors in a FIWARE-based solution can represent a sensor or an actuator. A sensor can report the state of the world around it, while an actuator can change the state of the system by responding to control signals coming from the CB~\cite{fiware_IoT-Sensors}. Therefore, with the FIWARE framework, it is possible to implement the four modules present in the sensor architecture since the properties of an actuator include the possibility to set the sensor configuration parameters based on commands received from a CB.

\begin{figure*}
	\centering
		\includegraphics[width=.75\linewidth]{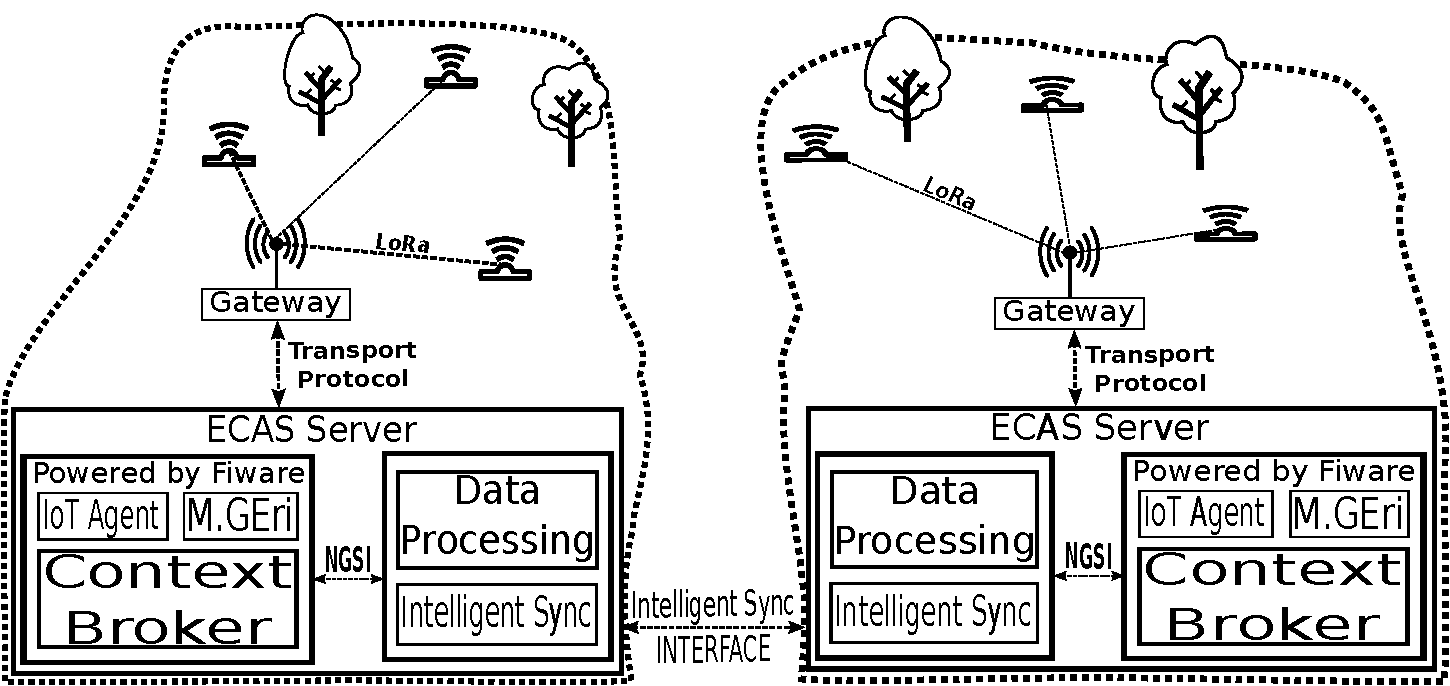}
	\caption{System Architecture ``Powered by FIWARE''.}
	\label{System_Architecture_PoweredbyFIWARE}
\end{figure*}

The ECAS server IoT Interface module can be represented by the IoT Agent. In addition to the IoT Agent, FIWARE provides a Monitoring GEri that aims to incorporate monitoring systems with the CB~\cite{fiware_Monitoring}. Thus, it is possible to use FIWARE to implement two other ECAS modules, Selective Bridge and Integration Interface. Since FIWARE can set configuration parameters of the sensors, it also enables the Sensor Manager module's implementation on the ECAS server's side. Any platform or solution built on top of FIWARE has only one mandatory component, a FIWARE Context Broker Generic Enabler~\cite{fiware_Catalogue}. Furthermore, since this solution has this Context Broker, it is named ``Powered by FIWARE''. Since this section suggests implementing ECAS using a CB, \textbf{Figure~\ref{System_Architecture_PoweredbyFIWARE}} shows how ECAS would be developed on top of the FIWARE framework.

The modules implemented by FIWARE are highlighted in \textbf{Figure~\ref{System_Architecture_PoweredbyFIWARE}} inside a box called ``Powered by FIWARE''. This box shows the three software components that make up the ECAS server, an IoT Agent, Monitoring GEri, and the CB. These components perform the functions of the Sensor Manager, Selective Bridge, Integration Interface, and IoT Interface modules. An NGSI interface handles the communication process between the ECAS modules and the modules implemented by FIWARE.

\textbf{Figure~\ref{System_Architecture_PoweredbyFIWARE}} represents the communication of sensors with the Gateway using LoRa, but other LPWAN technologies such as NB-IoT, LTE-M, and Weightless can be used for this communication. Likewise, the communication between the Gateway and the IoT Agent can use any of the supported transport protocols. \textbf{Figure~\ref{System_Architecture_PoweredbyFIWARE}} can represent different IoT Agent implementations. For example, it is possible to have an Ultralight-based IoT Agent in one area, while the other area might have a JSON-based IoT Agent, both running over the HTTP transport protocol. Examples of other supported transport protocols are AMQP and MQTT; in which case, it would be necessary to deploy an additional message  broker that is not represented in \textbf{Figure~\ref{System_Architecture_PoweredbyFIWARE}}. Among the ECAS servers, there is an interface called Intelligent Sync Interface, and this interface is responsible for synchronizing the data stored in each server and distributing it to the servers that may need this data.

According to the needs of each ECAS server, the CB processes and classifies the data collected from the context. Therefore, to implement the Intelligent Sync module it is possible to use artificial intelligence techniques to determine what should be shared and with which servers. The Intelligent Sync module is responsible for distributing the data to other ECAS servers on the battlefield. Instead of replicating every piece of data in all servers, some data distribution techniques already used in other works could enable the development of this module. Liu~\textit{et al.}~\cite{liu2020scalable} propose a distributed mechanism that works offline and online to determine which data center to place data based on read/write requests. The Intelligent Sync module can benefit from the scalability and distributed architecture introduced by this mechanism. In another work, Liu~\textit{et al.}~\cite{liu2018learning} proposed a framework called DataBot, which uses techniques such as neural networks (NN) and reinforcement learning (RL) to create data placement policies for a system with distributed data centers. The data that needs to be shared by ECAS servers involves information collected from the physical context. Therefore, implementing artificial intelligence and machine learning (ML) techniques to assist in the intelligent distribution of data can increase the system's efficiency.

\section{Experiments and Results}
\label{sec:4}
To demonstrate the efficiency of the proposed system, simulations have been performed as a proof-of-concept. The network simulator used was the \textit{``ns-3 - discrete-event network simulator''}. This simulator offers a LoRaWAN module, which allows the simulation of a sensor network based on LoRa technology. Also, the NYUSIM simulator~\cite{sun2017nyusim} was employed to calculate the attenuation in the wireless communication links by varying weather factors, as explained in detail in the following sections.

\subsection{Experimental Setup}
\label{sec:4.1}
\sloppy The implemented ECAS architecture uses LPWAN technology to communicate sensors distributed on the battlefield with servers. In the simulation developed as a proof-of-concept, the chosen LPWAN communication technology was LoRa. LoRa is a proprietary spread spectrum modulation technique that is derivative of chirp spread spectrum modulation (CSS); this technique has been used in military applications because of its long-range coverage and interference robustness \cite{sinha2017survey}. Although LoRa's characteristics make it suitable for the operation of the proposed system, factors such as support for applications with real-time requirements \cite{leonardi2019rt}\cite{khutsoane2019watergrid}, interoperability with command and tactical control systems \cite{jalaian2018evaluating} were also considered for the choice of this technology.

LoRa radio has four configuration parameters:  carrier frequency,  spreading factor (SF), bandwidth, and coding rate. Setting these parameters determines signal robustness, power consumption, and transmission range~\cite{bor2016lora}. An important feature of LoRa that was explored in this simulation was the Data Rate (DR). The DR consists of the combination of two LoRa configuration parameters: the bandwidth and the spreading factor. The highest DR in LoRa communication provides a greater volume of data transmitted over shorter distances. Although the amount of data transmitted is less in the lower DR, the signal strength is greater, allowing communication between Sensor and Gateway over long distances and unfavorable weather conditions. These factors are important to consider regarding the application scenario under concern.

\begin{figure}
	\centering
		\includegraphics[width=1\linewidth]{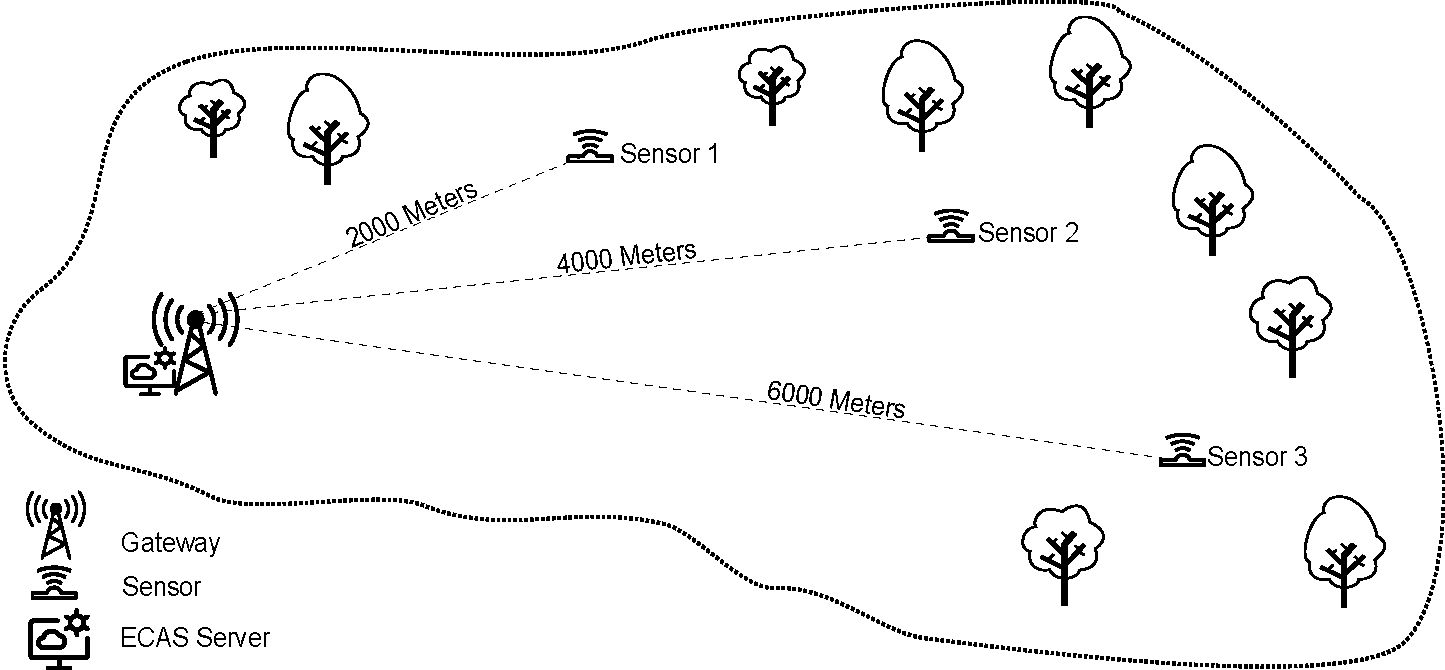}
	\caption{Scenario of the Experiment.}
	\label{Scenario_of_the_Experiment}
\end{figure}

An experiment with the following parameters was carried out. A LoRa gateway operating on three channels of 868MHz, one channel for each sensor, was installed at 15 meters height. Three sensors with a fixed location (without mobility) at 2,000, 4,000, and 6,000 meters away from the gateway, and at 1.7 meters height. Each round of the experiment represented 24 hours of simulation and each node sent a packet to the gateway periodically every 60 seconds. \textbf{Figure~\ref{Scenario_of_the_Experiment}} illustrates the scenario of the experiment just described.

At each round of the experiment, the DR of the sensors were modified. For each DR, an attenuation representing 1mm/h of rain was applied to the signal, then run one round. Therefore, 151 rounds of the experiment were executed for each DR. Starting, for each DR, in 0mm/h of rain, which represented a channel free of attenuation due to rain, and ending with 150mm/h, representing a heavy attenuation scenario.

In the first stage of the experiment, the same DR was configured in the three sensors, ranging from DR-0 to DR-5. Being DR-5 the highest packet rate and the lowest reach, and the DR-0 the lowest packet rate and greater reach of signal. \textbf{Table~\ref{DataRatesoftheExperiment}} presents the combinations of SF and bandwidth for the DRs applied during the experiment. The DR-5 represents the configuration with the ability to transmit the most massive volume of data. While the DR-0 represents a more robust signal to attenuation, therefore it reaches greater distances. Although the robustness of the signal experienced by DR-0 is better than the other DRs, the volume of data transmitted by it is lower than the higher DRs. Therefore, applying the appropriate DR to the sensors distributed across the battlefield, considering the physical and environmental context, increases the network's reliability by offering a communication infrastructure adaptable to the context.

\begin{table}[width=.5\linewidth,cols=3]
\caption{Data Rates of the Experiment.}\label{DataRatesoftheExperiment}
\begin{tabular*}{\tblwidth}{LLL}
\toprule
DR & SF & Bandwidth  \\
\midrule
5& 7& 125kHz  \\
4& 8& 125kHz  \\
3& 9& 125kHz  \\
2& 10& 125kHz \\
1& 11& 125kHz \\
0& 12& 125kHz \\
\bottomrule
\end{tabular*}
\end{table}

In the second stage of the experiment, the goal was to observe how ECAS acts in the monitored environment. In this approach, ECAS collects the battlefield information and adapts the DR of each sensor to the best DR according to the observed weather conditions. This adaptation is made using two approaches called: Conservative and Aggressive. In the Conservative approach, ECAS receives the rain rate in mm/h from each sensor, calculates the attenuation of this factor in the communication link, and, if necessary, reconfigures the DR of a given sensor to increase signal robustness or improve transmission rate. In this approach, the system modifies the sensor's DR before reaching a threshold. In contrast, in the Aggressive approach, ECAS expects a given sensor to decrease in the PDR, and only then it reconfigures that sensor's DR.

The Conservative approach has been configured to reduce the DR of the sensor by one, which can be affected by attenuation, 5mm/h of rain before the attenuation causes any impact on the PDR. Therefore, before any sensor has a reduction in PDR, its DR is reduced to increase signal robustness, although the packet sending rate also reduces. In the Aggressive approach, the highest DR is used until a reduction in PDR from a given sensor is observed. Only then adjustments are made to the DR of the respective sensor. This approach allows better use of higher data rates in scenarios with small climate variation.

\subsection{Results}
\label{sec:4.2}
\begin{figure}
	\centering
		\includegraphics[width=1\linewidth]{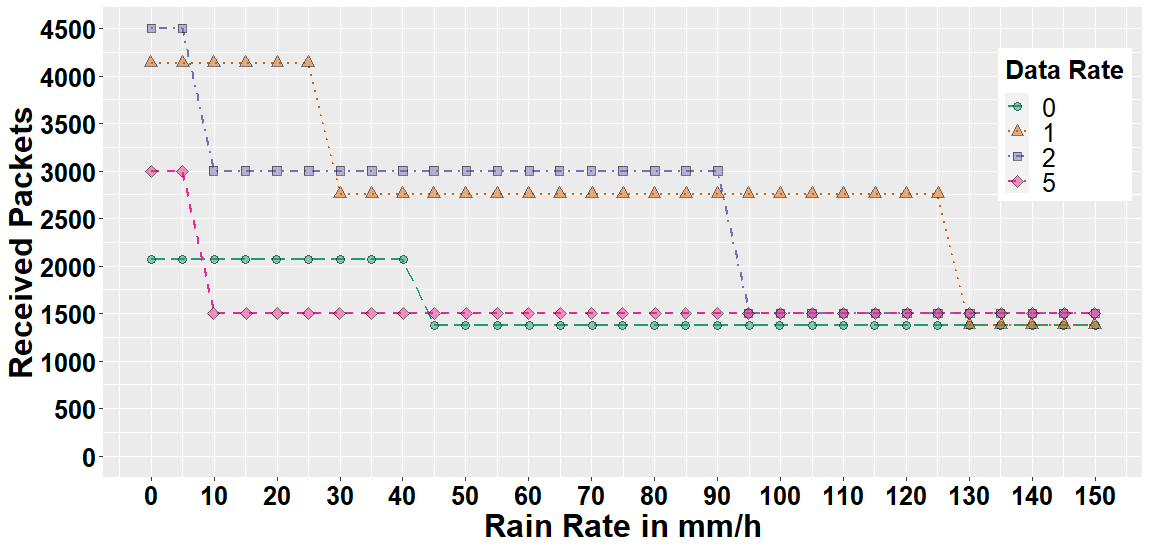}
	\caption{Delivered Packets Per Data Rate.}
	\label{Delivered_Packets_Per_Data_Rate}
\end{figure}

\textbf{Figure \ref{Delivered_Packets_Per_Data_Rate}} shows the results of the first stage of the experiment, in which a fixed DR on the three sensors were set. The Y-axis of \textbf{Figure~\ref{Delivered_Packets_Per_Data_Rate}} represents the received packets by the gateway per round of the experiment; the X-axis represents the increase of rain in mm/h that goes from 0mm/h to 150mm/h. For better visualization of the results, the DR-3 and DR-4 have been hidden. However, the information of these DRs can be found in \textbf{Table~\ref{Sent,Received,LostPacketsandPDRDataRate}}, which shows the number of packets sent, received, lost, and PDR in percentage for each DR applied in the experiment. During the experiment, it was observed that the packet sending rate experienced by the three sensors from DR-5 to DR-2 was the same, 1500 packages per experimental round. While for DR-1 and DR-0, all sensors suffered a fall in the packet sending rate. It reached approximately 1380 packets sent per round with DR-1 and 690 with DR-0 per sensor.

It is possible to observe in \textbf{Table~\ref{Sent,Received,LostPacketsandPDRDataRate}} that, because the packet sent rate experienced by the DRs from 5 to 2 is the same, for the parameters applied to the experiment, the use of DR-5 does not present the best performance compared to DR-4, DR-3, and DR-2. At the beginning of the experiment, sensors one and two, located at 2,000 and 4,000 meters away from the gateway, connect from 0mm/h to 5mm/h of rain. At the same time, the third sensor, located at 6,000 meters, has no connection since the beginning of the experiment. From 5mm/h on, sensor 2 loses connection to gateway due to the attenuation caused by the increase in rain rate. For DR-2, whose packet sent rate is the same as that experienced by DR-5 and the signal is more robust, at the beginning of the experiment, all nodes have a connection to the gateway, reaching a PDR of 1,500 per node (4,500 in total). This PDR is experienced up to 5mm/h of rain, when the PDR drops to 3,000 packets per round, 1,500 packets of each of sensors 1 and 2, and remains up to 90mm/h of rain when the second node loses connection to the gateway.  

\begin{table}[width=.9\linewidth,cols=5]
\caption{Sent, Received, Lost Packets and PDR\% per Data Rate.}\label{Sent,Received,LostPacketsandPDRDataRate}
\begin{tabular*}{\tblwidth}{LLLLL}
\toprule
DR & Sent & Received & Lost & PDR\% \\
\midrule
5& 139,500& 49,500& 90,000& 35\%  \\
4& 139,500& 57,000& 82,500& 41\%  \\
3& 139,500& 66,000& 73,500& 47\%  \\
2& 139,500& 78,000& 61,500& 56\%  \\
1& 128,216& 86,851& 41,365& 68\%  \\
0& 64,139& 48,959& 15,180& 76\%   \\
\bottomrule
\end{tabular*}
\end{table}

DR-0 and DR-1 achieved a better PDR, although the number of packets sent is lower than higher DRs. In DR-1, the packet sent rate per node approaches the 1,380 packets, while in DR-0, the rate drops to approximately 690. \textbf{Figure~\ref{Delivered_Packets_Per_Data_Rate}} shows that the DR that performed better in contrast to the others was DR-1. In DR-1, the packets sent rate comes near 1,500 packets, which is experienced by the higher DRs, and the signal's robustness makes all three sensors connect with gateway at higher rainfall rates. Using DR-1, the furthest sensor can connect with the gateway up to 25mm/h of rain. Using DR-2 and above, this sensor's connection does not exceed 5mm/h of rain. Sensor 2 can connect to the gateway up to 125mm/h of rain with DR-1, contrasting with the 90mm/h of the DR-2 and lower rainfall rates in the higher DRs. In the DR-0, which represents the most robust signal, but with the lowest packet sent rate, the farthest node had a connection with gateway up to 40mm/h of rain when it loses connection. At this DR configuration sensors 1 and 2 remain connected to the gateway up to 150mm/h of rain.

\begin{figure}
	\centering
		\includegraphics[width=1\linewidth]{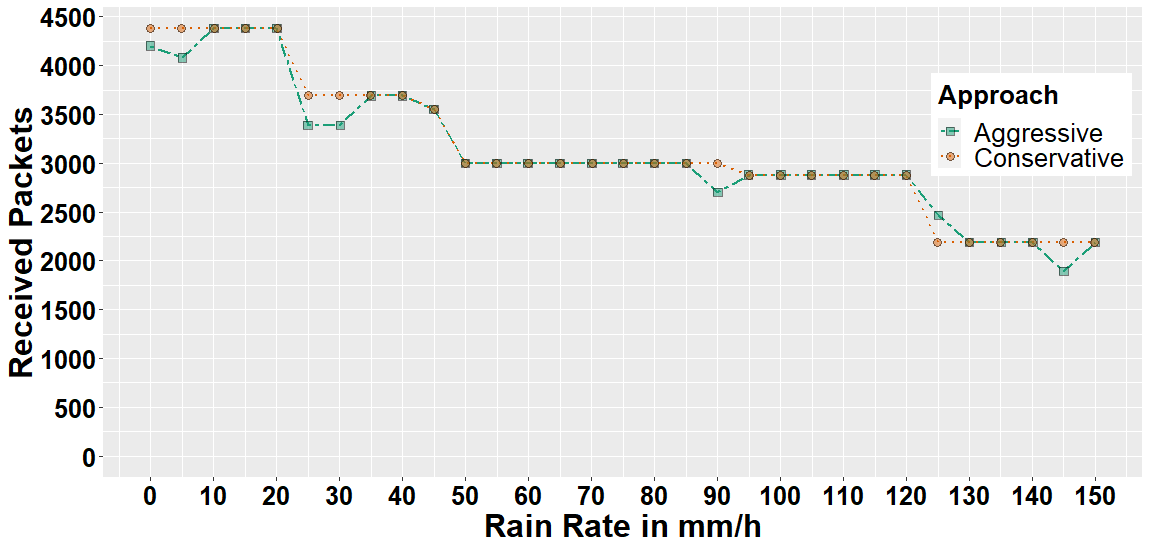}
	\caption{Delivered Packets Adaptive Aggressive X Conservative Approach.}
	\label{DeliveredPacketsAdaptiveAggressiveXConservativeApproach}
\end{figure}

\textbf{Figure~\ref{DeliveredPacketsAdaptiveAggressiveXConservativeApproach}} presents the results of the experiment considering a scenario with the implementation of the ECAS. The DRs adapted in this scenario, consider the rain rate and sensor's distance to the gateway. The Y-axis in the chart of \textbf{Figure~\ref{DeliveredPacketsAdaptiveAggressiveXConservativeApproach}} represents the received packets' rate, while the X-axis represents the rain rate in mm/h. For both the Aggressive adaptive approach and the Conservative adaptive approach, the results obtained were higher than fixed DRs.

In the Aggressive approach, each loss detection, the sensor DR was reduced by one to increase signal robustness and improve the gateway connection. This approach allows the use of DR with a higher rate still considering the climatic conditions at the time of transmission. The Conservative approach behaves differently. When a given rain level is observed, the DR of the sensor is decremented by one 5mm/h of rain before the signal can be affected. Therefore, the sensor should not lose connection to the gateway whatsoever. However, suppose the rain does not exceed this threshold of 5mm/h. In that case, the sensor will have stopped using the DR with a higher packet sending rate until the system sees a reduction in the rain rate, which would allow an increase in DR considering the new rain scenario.

In \textbf{Figure~\ref{DeliveredPacketsAdaptiveAggressiveXConservativeApproach}}, the dotted line represents the Aggressive approach. At the beginning of the experiment, this approach presents some received packet lower than the Conservative approach. This behavior occurs because the Aggressive approach tries to use the largest possible DR from the beginning of the experiment, then starts the experiment with losses until it finds the ideal DR. From the 10mm/h of rain, the Aggressive approach achieves the same number of packet received as the Conservative approach. During the experiment, every now and then, the Aggressive approach achieves a lower number of packet received than the Conservative approach; this is due to the method used to determine when to adjust the DR in the Aggressive approach. At 120mm/h, the Aggressive approach performs better than the Conservative approach. It was expected that this would happen at some moment. While the Conservative approach prevents losses, the Aggressive approach uses losses as a factor in determining when to adjust the DR. At this moment of the experiment, the loss occurred in the 124mm/h of rain, which ensured a number of packet received higher for the Aggressive approach in the range of 120mm/h to 124mm/h of rain.

The experiment, which served as a proof-of-concept to exemplify the functioning of the ECAS, adopted the premise that the rate of rain only increases. Thus, the experiment started with the rain rate at 0mm/h and ended in 150mm/h. This premise demonstrates what the system's behavior would be if adopted for the battlefield scenario where the rain only increases. However, in a real scenario, rain rates do not increase linearly. The rain can increase, stabilize, or even reduce before stopping. It is not possible to determine when one of these events will occur. Thus, two approaches have been proposed to demonstrate the behavior for the system considering the performance and reliability requirements.

\begin{figure}
	\centering
		\includegraphics[width=1\linewidth]{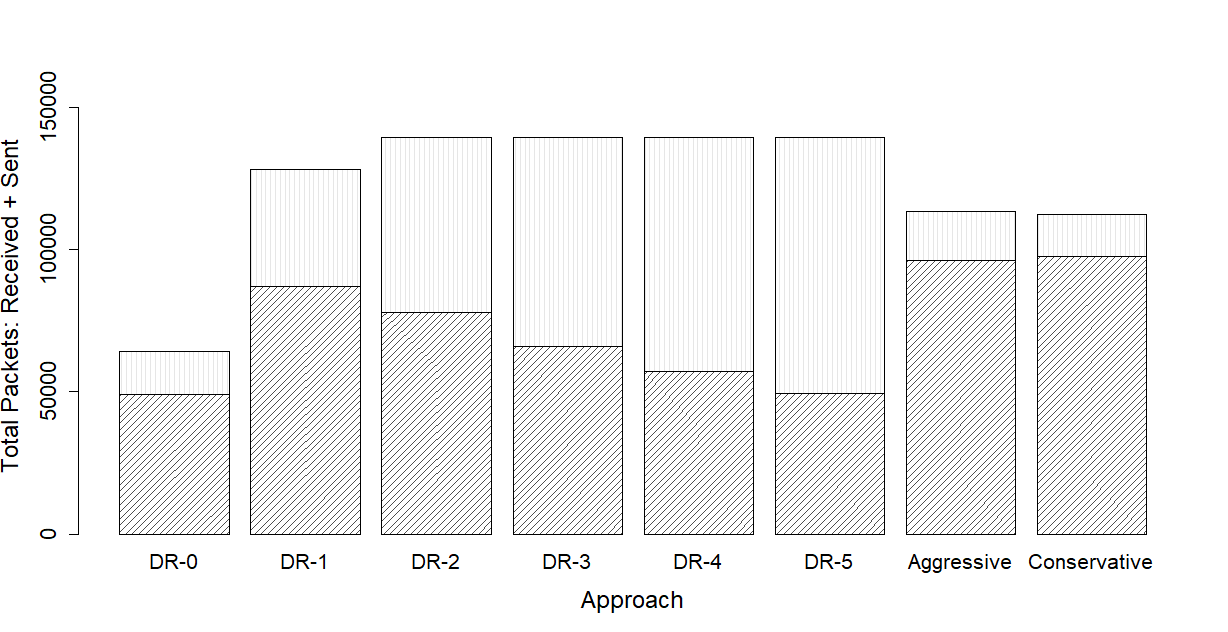}
	\caption{Total Packets Received plus Sent per Approach - Fixed Data Rate, Adaptive Aggressive and Conservative.}
	\label{TotalPackets}
\end{figure}

\textbf{Figure~\ref{TotalPackets}} presents a comparison of the experiment performed with fixed DR and the Aggressive and Conservative approaches of the ECAS. On the Y-axis of \textbf{Figure~\ref{TotalPackets}}, the Total Packets Received plus the Total Packets Sent is displayed. Each bar represents a setting applied in the experiment. By adopting any of the approaches offered by ECAS, it is possible to obtain a higher PDR than fixed DR approaches. The PDR experienced by DR-1 and DR-2 is one that is closer to the PDR experienced by the approaches offered in the ECAS. \textbf{Table~\ref{ResultperExperimentSetup}} presents the employed approach, the number of packets sent, and the PDR for each experiment round. Approaches using ECAS offer a packet sending rate that approximates the approaches with the best fixed DR performance. However, the amount of received packets is much higher when ECAS is in place to monitor and adapt to the environment. 

\begin{table}[width=.9\linewidth,cols=3]
\caption{Result per Experiment Setup.}\label{ResultperExperimentSetup}
\begin{tabular*}{\tblwidth}{LLL}
\toprule
Approach & Packets Sent & PDR\% \\
\midrule
Fixed DR-5& 139500& 35\%  \\
Fixed DR-4& 139500& 41\%  \\
Fixed DR-3& 139500& 47\%  \\
Fixed DR-2& 139500& 56\%  \\
Fixed DR-1& 128216& 68\%  \\
Fixed DR-0& 64139& 76\%  \\
ECAS-Aggressive& 113309& 85\%  \\
ECAS-Conservative& 112231& 87\%  \\
\bottomrule
\end{tabular*}

\end{table}

\section{Conclusion}
\label{sec:5}
The unpredictability of the battlefield context makes it necessary to implement systems that improve the decision-making process time-effectively. The present work proposed a context-aware monitoring system to adapt network configuration parameters to changes in the climatic factors observed on the battlefield. A simulation was developed as a proof-of-concept to demonstrate how the proposed approach performs under varied weather conditions. 

The simulation results showed that the dynamic adaptation of network parameters can make the communication process more efficient. Comparing the best results of the fixed and adaptive approaches, it was achieved an 11\% increase in the total packets delivered, while simultaneously reducing the total packets sent by 12\%. From these results, it is possible to infer that the use of the adaptive approach, in addition to making the communication process more efficient, also has a potential to reduce resource/energy consumption.

Several works report the issue of interference caused by environmental factors in wireless network technologies \cite{anastasi2004performance}\cite{federici2016review}\cite{rangarajan2015evaluating}\cite{bruzgiene2020quality}. Although the LoRa technology was used to develop a proof-of-concept in this work, the proposed system is technology agnostic. Therefore, in addition to the feasibility of implementing ECAS in an existing sensor network, it is possible to use the system to assist in configuring the parameters of various wireless network technologies.

Issues related to security such as confidentiality, integrity, and authenticity, despite their importance in military networks, were left outside the scope of this work. However, the technologies used in this work already address these issues in other layers that can be implemented to provide security for the proposed system \cite{alliance2017lorawan}.

As future work, the  plan is to implement ECAS on existing IoT platforms like FIWARE, as presented in \textbf{Session~\ref{sec:implementation}}. Another factor to be evaluated in the system is the impact of temperature on signal quality. The ECAS should adapt the network configuration parameters analyzing multiple climatic factors, besides the rainfall attenuation considered in this work. Finally, ECAS will be implement using real-world devices to validate in practice the system's efficiency in terms of resilience, network survivability, as well as energy consumption.

\section*{Acknowledgments}
\label{sec:6}

This study was partially funded by CAPES - Finance Code 001. We also thank the funding of CNPq, Research Productivity Scholarship grants ref. 313893/2018-7, ref. 309505/2020-8 and Universal research project ref. 420109/2018-8. We also thank to the Brazilian Army for the support provided via the research project S2C2, ref. 2904/20. This work was also supported by the São Paulo Research Foundation (FAPESP) grant \# 2020/05182-3.

\printcredits

\bibliographystyle{cas-model2-names}

\bibliography{cas-refs}

\newpage
\bio{figs/Guilherme_3x4.png}
Guilherme Rotth Zibetti (guilherme@ufrgs.br) is an M.Sc. student in computer networks at the Federal University of Rio Grande do Sul (UFRGS), Brazil. He holds a degree in Computer Networks from Faculdade Senac Porto Alegre, Brazil, 2019. His research interests are currently related to wireless networks, sensor networks, software-defined networks, ad hoc network, and the Internet of Things.

\endbio

\bio{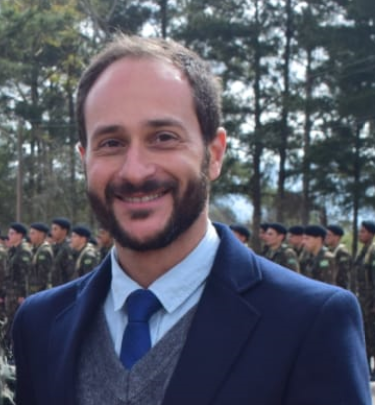}
Edison Pignaton de Freitas is a computer engineer graduated from the Military Institute of Engineering in Rio de Janeiro, Brazil (2003) He received an M.Sc. degree in computer science from the Federal University of Rio Grande do Sul (UFRGS), Porto Alegre, Brazil (2007), and the Ph.D. degree in computer science and engineering in the area of wireless sensor networks from Halmstad University, Halmstad, Sweden (2011). Currently, he holds an Associate Professor with UFRGS, affiliated to the Graduate Programs in Computer Science and Electrical Engineering,  acting mainly in the following areas: Wireless sensor networks,  real-time and embedded systems, avionics, and unmanned vehicles.
\endbio

\bio{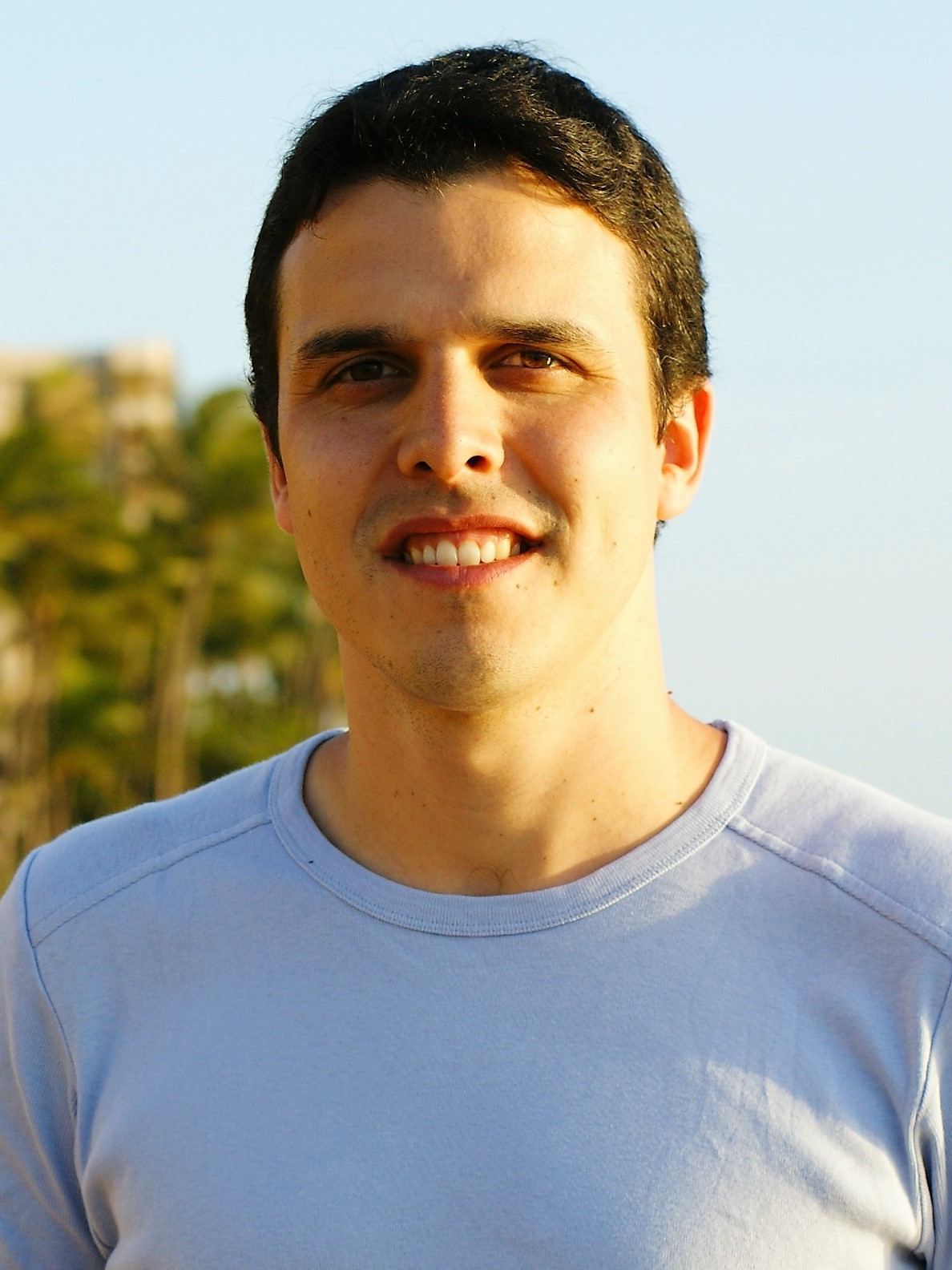}
Juliano Araujo Wickboldt is an associate professor at the Federal University of Rio Grande do Sul (UFRGS) in Brazil. He holds both M.Sc. (2010) and Ph.D. (2015) degrees in computer science from UFRGS. Juliano was an intern at NEC Labs Europe in Heidelberg, Germany for one year between 2011 and 2012. In 2015, Juliano was a visiting researcher at the Waterford Institute of Technology in Ireland. His research interests include softwarized networking, IoT, and 5G technologies.
\endbio

\end{document}